\documentclass[structabstract]{aa} 
\usepackage{epsfig}
\usepackage{graphicx}
\usepackage{wasysym}
\usepackage{amssymb}
\usepackage{stmaryrd}
\usepackage{rotating}
\usepackage{longtable}
\usepackage{lscape}
\usepackage[normalem]{ulem}

\usepackage[usenames,dvipsnames]{xcolor}

\usepackage{graphicx}
\usepackage{txfonts}
\begin{document}

   \title{Study of open clusters within 1.8 kpc and understanding the Galactic structure}

   \author{Y.~C.~Joshi$^{1}$\thanks{E-mail: yogesh@aries.res.in}\thanks{Data used in the present study is only available in the electronic form at the CDS via anonymous ftp to cdsarc.u-strasbg.fr},
A. K. Dambis$^{2}$,
A. K. Pandey$^{1}$,
S. Joshi$^{1}$
          }
   \institute{
$^{1}$Aryabhatta Research Institute of Observational Sciences (ARIES), Manora peak, Nainital 263002 India\\
$^{2}$Sternberg Astronomical Institute, Lomonosov Moscow State University, Universitetskii pr. 13, Moscow 119992, Russia          
}
   \date{Received 17 May 2016 / accepted 20 June 2016}

\abstract
{Significant progress has been made in recent years to understand the formation and evolution of our Galaxy, but we still lack a complete understanding of the Galaxy and its structure.}
{Using an almost complete sample of Galactic open star clusters within 1.8 kpc, we aim to understand the general properties of the open cluster system in the Galaxy and probe the Galactic structure.} 
{We first extracted 1241 open clusters within 1.8 kpc of the Sun from the Milky Way Star Clusters (MWSC) catalog. Considering it an almost complete sample of clusters within this distance, we performed a comprehensive statistical analysis of various cluster parameters such as spatial position, age, size, mass, and extinction.}
{We find an average cluster scale height of $z_h = 60\pm2$ pc for clusters younger than 700 Myr, which increases to $64\pm2$ pc when we include all the clusters. The $z_h$ is found to be strongly dependent on $R_{GC}$ and age, and on an average, $z_h$ is more than twice as large as in the outer region than in the inner region of the solar circle, except for the youngest population of clusters. The solar offset is found to be $6.2\pm1.1$ pc above the formal Galactic plane. We derive a local mass density of $\rho_0$~=~0.090~$\pm$~0.005~$M_{\odot}/{\rm pc}^3$ and estimate a negligibly small amount of dark matter in the solar neighborhood. The reddening in the direction of clusters suggests a strong correlation with their vertical distance from the Galactic plane with a respective slope of $dE(B-V)/dz$ = 0.40$\pm$0.04 and 0.42$\pm$0.05 mag/kpc below and above the GP. We observe a linear mass-radius and mass-age relations in the open clusters and derive the slopes of $dR/d(logM)~=~2.08\pm0.10$ and $d(logM)/d(logT)~=~-0.36\pm0.05$, respectively.}
{The dependence of the spatial distribution of clusters on their age points to a complex interplay between cluster formation and survivability within the Galaxy. The geometrical characteristics of a significant number of clusters enabled us to understand large-scale spatial properties of the cluster systems within the Galaxy. The structural and physical parameters of clusters allowed us to check mutual correlations between the individual parameters.}

\authorrunning{Y. C. Joshi et al.}

\titlerunning{Study of open star clusters within 1.8 kpc}

\maketitle
\section{Introduction}
The solar system is located within the Galactic disk itself, therefore studying the structure of our Galaxy is a challenging task. Although considerable progress has been achieved in recent times to understand the formation and evolution of the Galaxy, we still lack a comprehensive understanding of the subject. One of the ways to investigate the Galactic structure is through studying the properties of various classes of objects that populate the Galaxy. For example, the formation and evolution of the Galaxy can be probed through open star clusters, which are groups of stars that formed from the same molecular cloud and have roughly the same age, distance, and chemical composition and are visible up to large distances. Since open clusters are distributed throughout the Galactic disk and span a wide range in age, they are a useful tool for studying the effects of the dynamical evolution of the Galactic disk and trace the spiral arms (Carraro et al. 1998; Chen et al. 2003; Piskunov et al. 2006). In addition, these objects are excellent tracers of the chemical evolution of the Galactic disk because they are among the best candidates for providing precise information on both the ages and chemical compositions at various positions in the Galaxy and up to a large Galactocentric distance (Friel et al. 2002, Yong et al. 2012, Frinchaboy et al. 2013).

To understand the Galactic structure, a statistical analysis of a large number of star clusters in the Galaxy is very important because individual values of cluster parameters may be affected by the considerable uncertainty in their determination. Such studies have been carried out by many authors in the past. Lyng\.{a} (1982) has published a catalog of Galactic open clusters, and  Lyng\.{a} \& Palous (1987), Pandey et al. (1987), Janes et al. (1988), and others used these data to discuss some of the general properties of the open cluster system and the Galaxy. Since then, understanding of the Galactic structure has become an incremental process with the continuous addition of new data on star clusters and their improved parameters. In recent years, the number of observed open clusters has grown rapidly, which has paved the way for understanding the Galactic structure in greater detail. Studies like this have already been carried out by many authors on the basis of the DAML02 open cluster catalog (Dias et al. 2002), the sample of which is updated with recently published data on the star clusters\footnote{The most recent version of  DAML02 catalog released on 28 January 2016 lists more than 2100 open clusters.}. Over the past ten years, Tadross et al. (2002), Joshi (2005), Bonatto et al. (2006), Bonatto \& Bica (2011), Santos-Silva \& Gregorio-Hetemand (2012), Buckner \& Froebrich (2014), and many others have used this catalog to study the Galactic structure in great detail. However, the Dias catalog is heterogeneous in nature because a variety of techniques and observational setups were used by different authors to derive the cluster parameters. Recently, Kharchenko et al. (2013) have published a large catalog within their Milky Way Star Clusters (MWSC) project, for which the PPMXL (R\"{o}ser et al. 2010) and 2MASS (Skrutskie et al. 2006) all-sky catalogs were used to determine various cluster parameters. The MWSC is believed to be homogeneous because the parameters of all the clusters are estimated through the same method and uncertainties in the obtained parameters are also systematic. It is extremely important to have cluster parameters derived in a homogeneous manner to carry out a reliable study of the Galactic cluster population. It is therefore imperative to re-investigate the Galactic structure with this large sample of data.

In the present study, we aim to carry out a comprehensive statistical analysis of various parameters of the open star clusters and examine correlations between different parameters such as age, size, mass, extinction, and their positions in the Galaxy to understand the formation history of star clusters, their properties, and the Galactic structure. The organization of the paper is as follows. The detail of the data is explained in Sect. 2. In Sect. 3 we describe the completeness of the cluster sample. In Sect. 4 we examine the spatial distribution of clusters in the Galaxy. In Sect. 5 we analyze the correlation between different parameters. The main results of the study are summarized in Sect. 6.
\section{Data}\label{data}
We compiled a catalog for our analysis from the MWSC catalog that gives the physical parameters of 3006, 139, and 63 star clusters in the series of papers published by Kharchenko et al. (2013), Schmeja et al. (2014), and Scholz et al. (2015), respectively. We first merged these three samples and then selected only open clusters after excluding the entities marked as globular clusters, dubious clusters, and asterisms because the properties of these populations are quite different from those of the open clusters. Finally, we were left with 2858 open clusters (hereafter we only use the term ``cluster" in place of ``open cluster" except when
stated otherwise). It should be noted here that the MWSC survey is dominated by older clusters, which is attributed to the NIR basis of the survey (Kharchenko et al. 2013). The masses of 650 clusters were taken from Piskunov et al. (2007, 2008). We also used the rectangular coordinate system ($X$, $Y$, $Z$), which is defined as 
X = d~cos(b)~cos(l); Y = d~cos(b)~sin(l); Z = d~sin(b). The Galactocentric distance $R_{GC}$ of the cluster is determined by the following relation
$$
R_{GC} = \sqrt{R_\odot^{2}+ (d~cos(b))^{2}-2 R_\odot~d~cos(l)~cos(b),}
$$
where the $X$, $Y$, and $Z$ axes point toward the Galactic center, to the direction of Galactic rotation, and to the North Galactic pole, respectively. Here, $d$, $l,$ and $b$ are the heliocentric distance, Galactic longitude, and Galactic latitude, respectively. The distance of the Sun from the Galactic center, $R_\odot$, is taken as 8 kpc (e.g., Reid 1993, Malkin 2012, Honma et al. 2012), which is slightly lower than the IAU recommended value of 8.5 kpc (Kerr \& Lynden-Bell 1986).

To examine various cluster parameters as a function of their ages, we subdivided the cluster sample into three different populations, namely young, intermediate, and old clusters. Young clusters are the objects that probe the spiral structure and thin disk, an inner layer of the Galactic disk where most of the dust and gas lie. Dias \& L{\'e}pine (2005) and Moitinho (2010) suggested that young clusters up to $\sim$ 20 Myr remain in their parent arms and later start drifting away from the spiral arms and fill the inter-arm regions. We therefore considered objects with ages of up to 20 Myr as young clusters. For the intermediate-to-old age boundary, the Hyades cluster is generally assumed to be an ideal candidate. It has an age of about 700 Myr (Friel 1999). Hence we considered objects with ages from 20 to 700 Myr as intermediate-age clusters, while those with ages greater than 700 Myr are old-age clusters. In the following analysis, we use the nomenclature YOC, IOC, and OOC for the young, intermediate, and old age open clusters, respectively. We note that the age boundaries are not based on a uniform division of data but only approximately define whether a cluster is young, intermediate, or old.
\section{Observational selection effect and data completeness}\label{s_num_density}
It is a well-known fact that various observational selection factors such as distance and/or reddening effect play a significant role on the detectability of clusters and hence affect the observed distribution of clusters in the sky. The distant clusters that we see in the sky are probably either young clusters with extremely bright O and B stars or old clusters containing stars that have already left the main sequence and have become giants. Most of very young clusters contain significant numbers of luminous OB stars and are therefore observable at large distances against the background of field stars. However, those clusters that are deeply embedded in regions of dust and gas are still poorly observable at very distant regions. On the other hand, all the dust and gas have been used up in older clusters. They also host a sizable number of bright giant stars. Moreover, they are located at higher latitudes or in the Galactic anticenter direction where interstellar extinction is smaller. Therefore, when we look far beyond the solar position, the possibility of finding older clusters is higher than finding their relatively younger counterparts. However, when they become extremely old ($logT\gtrsim9.5$), most of the stars in some of these clusters are dissipated through dynamical processes. This reduces their overall luminosity and hence visibility at  larger distances. For intermediate-age clusters the lack of massive O and B stars, which quickly vanish with time, or the lack of any significant number of red giants reduces the overall cluster luminosity. Therefore they are hard to observe at large distances in the Galaxy. Clusters too close to the Sun are also difficult to identify because they are too sparse to be observed as dense regions. Furthermore, a significant number of clusters are difficult to observe at shorter wavelengths because of the high extinction near the Galactic plane (hereafter, GP). In general, the cluster detection rate depends upon several factors such as the number of stars, apparent magnitude at the main-sequence turn-off, size, heliocentric distance, stellar background projected onto the cluster direction, and detection limit of the survey. Various studies have proposed that the currently known clusters may be only 1 - 2\% of the actual number of total clusters present in the Galaxy (Lada \& Lada 2003, Chen et al. 2004, Piskunov et al. 2006, R\"{o}ser et al. 2010).
 
\begin{figure*}
\begin{center}
\includegraphics[width=16.0cm,height=16.0cm]{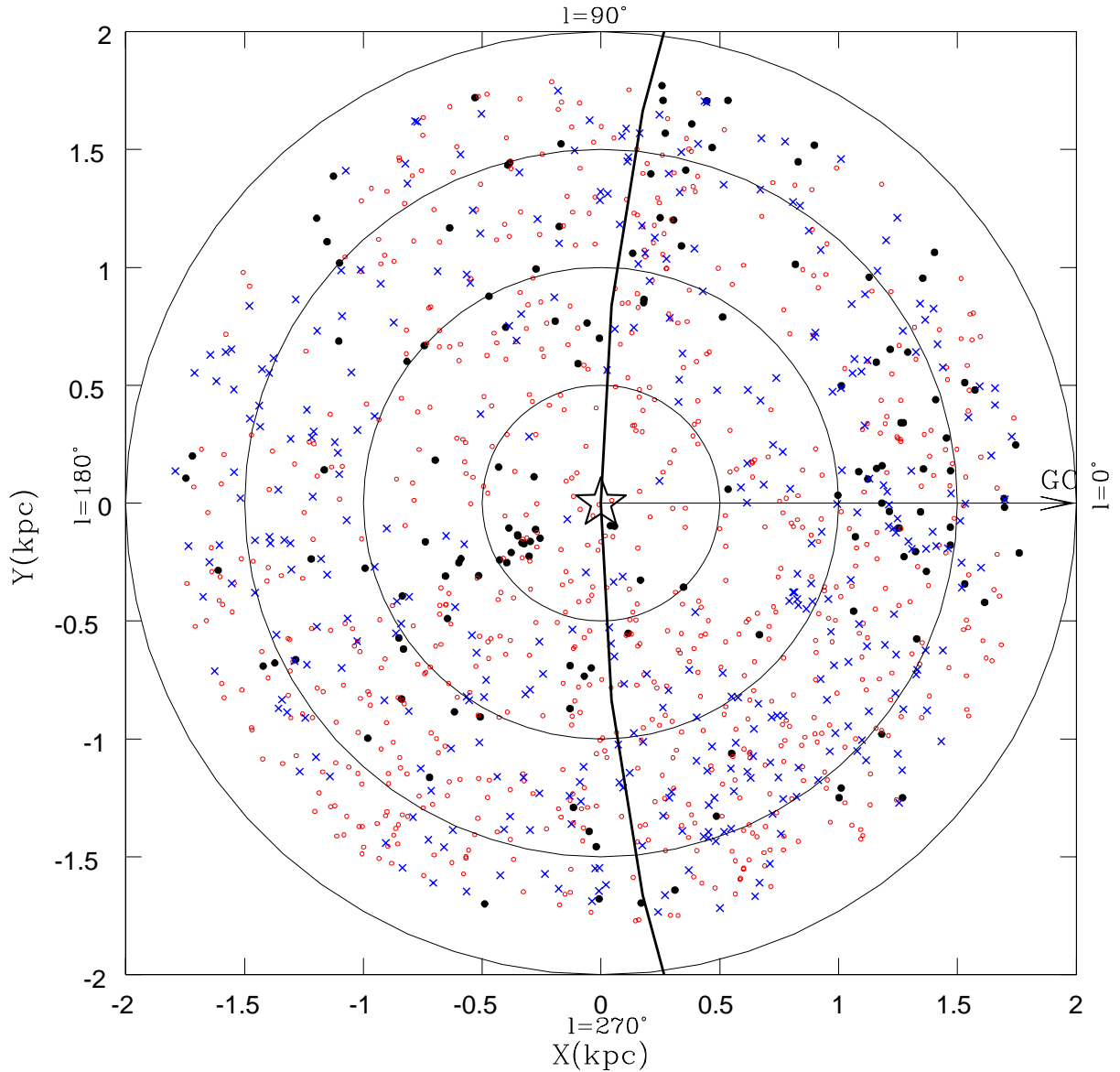}
\caption{Cluster distribution in the X-Y plane as projected on the GP. The filled circles, open circles, and crosses represent YOCs, IOCs, and OOCs, respectively. The asterisk at the center shows the position of the Sun, and concentric circles are drawn at an equal distance of 500 pc. The large dark circle represents the solar circle with a radius of 8 kpc, and the arrow indicates the direction to the Galactic center.}
\label{spatial}
\end{center}
\end{figure*}

To examine the completeness of the MWSC catalog, Kharchenko et al. (2013) determined the mean surface density for different ranges of heliocentric distances. Assuming a uniform density model for the distribution of clusters, they found their sample to be almost complete up to a distance $d$ = 1.8 kpc from the Sun for clusters of all ages except possibly the older clusters ($>$ 1 Gyr) and nearby clusters within about 1 kpc (Scholz et al. 2015). Most of the previous studies on the cluster distribution also pointed out that the known cluster sample may be complete within 1.6-1.8 kpc from the Sun (Joshi 2005, Bonnato \& Bica 2006, Buckner \& Froebrich 2014), although the present sample of clusters is the largest of all the studies made so far. To avoid any misinterpretation based on an incomplete set of data, we considered only clusters within 1.8 kpc in the present study. We found 1241 clusters in the MWSC catalog within 1.8 kpc of the Sun, of which 139, 750, and 352 clusters belong to YOC, IOC, and OOC populations, respectively. The catalog we used is available in electronic form at the CDS. 
\section{Spatial distribution of clusters}\label{s_spatial}
The spatial distribution of clusters provides a useful probe of the structure of the Galactic disk -- the younger systems delineate the spiral structure of the Galaxy, while the older ones, which are basically remnants of their larger initial populations, trace the kinematics of the outer Galaxy. To analyze the spatial distribution of clusters in the Galaxy, the distribution of the clusters projected onto the X-Y plane is  illustrated in Fig.~\ref{spatial}. We also draw the solar circle and four concentric circles around the solar position at 500 pc apart. Clusters defined as YOCs, IOCs, and OOCs  are shown by filled circles, open circles, and crosses, respectively. Clusters are found to be non-uniformly distributed in the sky as well as along the heliocentric distances. 

\begin{table*}
\centering
\caption{The number distribution of clusters within different heliocentric distance intervals. In the bracket, we give the surface density of cluster populations $\Sigma_{xy}$ (kpc$^{-2}$) within the given distance scale.}
\vspace{0.3cm}
\label{d_xy}
\begin{tabular}{c|ccccc} \hline
Cluster population  & \multicolumn{5}{c}{Number of clusters} \\
          & ~~ $<0.5$ kpc ~~ & ~~ 0.5-1.0 kpc ~~  & ~~ 1.0-1.5 kpc ~~   & ~~ 1.5-1.8 kpc ~~ & ~~~~~ up to 1.8 kpc ~~ \\ \hline \\
YOCs       & 18 (22.9) &  26 (11.0) &  58 (14.8) &  37 (11.9) & 139 (13.7)\\
IOCs       & 67 (85.3) & 189 (80.2) & 298 (75.9) & 196 (63.0) & 750 (73.7)\\
OOCs       &  6 ( 7.6) &  68 (28.9) & 169 (43.0) & 109 (35.0) & 352 (34.6)\\ \\
All        & 91 (115.9)& 283 (120.1)& 525 (133.7)& 342 (109.9)& 1241 (121.9) \\ \\
\hline
\end{tabular}
\end{table*}

In Table~\ref{d_xy} we list the number distribution of different cluster populations in four distance intervals. We found that IOCs are the dominant population of all the cluster populations in any distance interval. However, they are most dominant in the immediate solar neighborhood. We also determined the surface density of the cluster populations $\Sigma_{xy}$ (kpc$^{-2}$) in each distance interval, and the resulting values are provided in the brackets in Table~\ref{d_xy}. We found that the surface density of YOCs and IOCs is highest in the immediate solar neighborhood, while it decreases farther away from the solar position. For the OOCs the surface density is quite low, but it increases with distance from the solar position. When we considered all the clusters at any given distance interval, we found that the surface density of star clusters is highest in the 1.0-1.5 kpc interval. Combing all the clusters within 1.8 kpc of the Sun, we estimate an average surface number density of the clusters of 121.9 kpc$^{-2}$.

We note a huge deficiency of OOCs within 500 pc from the Sun, with only 6 out of a total of 91 clusters (6.6\%) found within this distance. This is about 1.7\% of the total OOCs found in our sample, and the remaining 98.3\% OOCs lie beyond 500 pc from the Sun. This tendency has also been reported in many past studies (e.g., Oort 1958), but it has become more evident with the addition of new clusters discovered in recent years. It is still unclear why so few OOCs are found in the solar neighborhood, but the prime reason seems to be that clusters are destroyed before they become too old, although the process depends upon the initial mass and density. Bergond et al. (2001) has estimated the destruction timescale to be about 600 Myr for clusters in the solar neighborhood. Lamers \& Gieles (2006) suggested that the dissolution of clusters in the solar neighborhood is primarily caused by four effects: stellar evolution, tidal stripping, spiral arm shocks, and encounters with molecular clouds, with the latter playing the dominant role. Lamers et al. (2005) calculated the dissolution time to be $t_{dis} = t_4 (M_i/10^4 M_\odot)^{0.62}$, where $t_4 = 1.3\pm0.5$ Gyr for clusters having $10^2 < M/M_\odot < 10^4$. For a typical 4000 $M_\odot$ cluster in the solar neighborhood, the dissolution time is about 700 Myr. This indicates that most of the massive clusters would dissolve in the solar neighborhood even before they reach the age of 1 Gyr, and this explains why there are so few OOCs present in the vicinity of the Sun.
\subsection{Distribution of open clusters in longitude}\label{ss_longitude}
\begin{figure}
\begin{center}
\includegraphics[width=8.5cm,height=5.5cm]{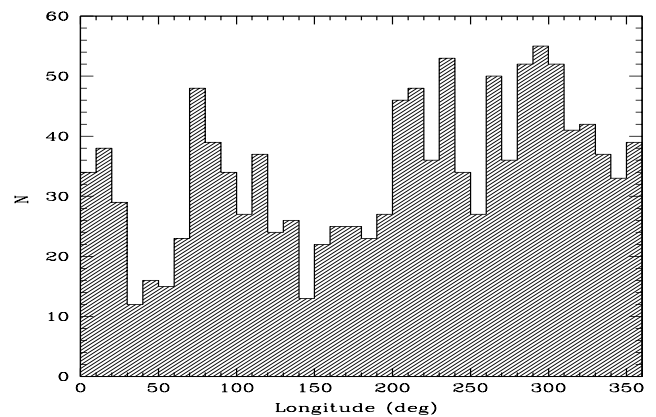}
\caption{Distribution of clusters along the Galactic longitude. The data are binned in 10-degree longitude intervals.}
\label{spa_distri_long}
\end{center}
\end{figure}

Figure~\ref{spa_distri_long} shows the longitude distribution of clusters as a histogram with a bin size of 10 deg. The extinction produced by dense interstellar material, combined with such factors as crowded stellar background and high gas density, leads to the low detection rate of clusters in the longitude range $320<l<60$ deg. The distribution shows maxima in regions of active star formation at  longitudes around 125 deg (Cassiopeia), 210 deg (Monoceros), 240 deg (Canis Major), and around 285 deg (Carina). It may well be that a considerable fraction of clusters is not identified because of the high extinction in certain regions in the sky. We observed deep minima primarily at two longitudes, one at around 50 deg (Sagitta) and another at around 150 deg (Perseus). The question is whether the massive molecular clouds are responsible for hiding the clusters in these directions, or if early destruction of the clusters has produced these dips. When we individually checked all the cluster samples including YOCs, IOCs, and OOCs in the longitude interval l $\sim$ 20-55 deg, we found the scarcity of all types of clusters in this Carina-Sagittarius arm, which clearly suggests that massive molecular clouds are indeed located in this direction. The second prominent dip is seen between 140 to 180 deg. Benjamin (2008) noted a similar effect in the distribution of stars with a gap in the Perseus Spiral Arm. Froebrich et al. (2010) inspected  the all-sky extinction maps of Rowels and Froebrich (2009), but found no signature of high-extinction molecular clouds  that might have prevented the cluster detection in these regions. 

Based on the reddening analysis of 722 clusters, Joshi (2005) has shown that the distribution of interstellar material shows a rather sinusoidal pattern along the Galactic longitude, and maximum and minimum reddening lie around $l\sim$ 50 and 230 deg, respectively. The distribution of clusters along the Galactic longitude, as shown in Fig.~\ref{spa_distri_long}, clearly matches these regions with relatively lower and higher number of clusters, suggesting that the cluster distribution is indeed affected by the distribution of interstellar material around the GP.
%
%

%
\subsection{Distribution of open clusters in the Galactic plane}\label{ss_GP}
\begin{figure}
\begin{center}
\includegraphics[width=8.5cm,height=5.5cm]{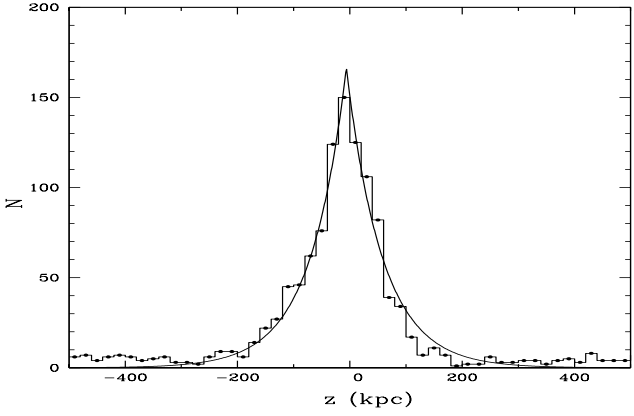}
\caption{Distribution of clusters perpendicular to the GP with a bin size of 20 pc in $z$. For clarity, the histogram is restricted to clusters within 0.5 kpc of the GP. The thick continuous line represents the best-fit exponential profile.}
\label{z_hist}
\end{center}
\end{figure}

Clusters are strongly concentrated around the GP because they are believed to be born from the interstellar material distributed in the form of a thin layer in the GP. In Fig.~\ref{z_hist} we show the frequency distribution of the clusters as a function of $z$. More than half of the clusters are found within $|z| < 50$ pc. The number density of the clusters in the Galactic disk is commonly expressed in the form of scale height in vertical direction, which is explained below.
%
%
\subsubsection{Scale height}\label{ss_zh}
The observed distribution of clusters perpendicular to the GP allows, in principle, determining the average velocity dispersion and scale height of the self-gravitating matter in the disk. This can be done with a fit to the observed distribution with the function of the following nature as given by Korchagin et al. (2003),
$$
  N (z) = N_0 sech^{\gamma}((z-z_0)/z_h), 
$$
where $N(z)$ is the number density of clusters normal to the GP, $N_0$ is the central cluster density, $z_0$ is the vertical shift about the Galactic symmetry plane, and  $\gamma = 2 \sigma_g^2/\sigma_{clus}^2$ is the ratio of the velocity dispersion of gravitating matter in the Galactic disk and that of the cluster sample, respectively. However, at $|z-z_0|>z_h$ , the above distribution becomes a decaying exponential law with respect to the GP in the form of
$$
N(z) = N_0~exp \left(- \frac{|z-z_0|}{z_h}\right). 
$$
We fit the above exponential profile to estimate the scale height of the distribution of clusters as $z_h = 64\pm2$ pc. The resulting best-fit exponential profile is represented by the thick continuous line in Fig.~\ref{z_hist}. We note, however, that the high value of $z_h$ is mainly influenced by the older sample of clusters. These older clusteres are primarily located at higher latitudes,
which results in a shallower wing in the $z$ distribution. If we exclude old clusters with ages $T>0.7$ Gyr, we infer a smaller scale height of $z_h = 60\pm2$ pc. A range in the cluster scale height has been reported by various authors who considered cluster samples of different ages and different distance scales. For example, Bonatto et al. (2006) derived $z_h = 48 \pm 3$ pc using clusters younger than 200 Myr and $57\pm3$ pc using 654 clusters of all ages. Piskunov et al. (2006) obtained $z_h = 56 \pm 3$ pc using clusters within 850 pc from the Sun. Joshi (2007) determined $z_h = 54 \pm 4$ pc using clusters younger than 300 Myr and lying within 250 pc of the GP. Considering clusters younger than 50 Myr, Zhu (2009) obtained $z_h = 51 \pm 5$ pc. Based on all these previous studies, it is conceived that the difference in measurements of scale heights is influenced by the selection criteria of the cluster sample.  To understand this problem in some detail, we study the variation of $z_h$ at different ages and $R_{GC}$ in our data.
\subsubsection{Dependence of $z_h$ on age}\label{ss_zh_age}
\begin{figure}
\begin{center}
\includegraphics[width=8.5cm,height=5.5cm]{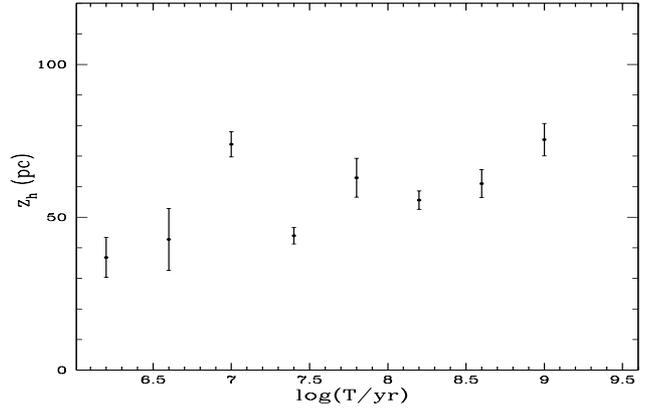}
\caption{Variation in cluster scale height as a function of $logT$.}
\label{Age_Zh}
\end{center}
\end{figure}

The precise determination of the disk scale height, $z_h$, is only possible when a significant number of clusters is available. To investigate the evolution of $z_h$ with cluster age, we need to have enough clusters between successive age bins. Considering the large sample of clusters available in the present catalog, we therefore investigated the evolution of $z_h$ with cluster age. We determined $z_h$ in $\Delta~logT=0.4$ bins to have a
sufficient number of clusters in each bin, but we did not find enough clusters beyond $logT>9.2$ where any reliable $z_h$ could be estimated. The resulting variation in $z_h$ with $logT$ is shown in Fig.~\ref{Age_Zh}, where the x-axis is the middle point of the $logT$ range in each bin. We found that $z_h$ is smallest for the YOCs and largest for the OOCs. An analysis by Joshi (2007) gives the scale height of $61\pm3$ pc for the OB stars that have a similar lifetime to that of the YOCs. Since clusters form close to the mid-plane, like  OB stars, their scale heights are comparable. The scale height for the clusters at a mean age of about 1 Gyr reaches about 75$\pm$5 pc, which is greater than the scale height of their younger counterparts. Buckner \& Froebrich (2014) also found the scale height to increase from about 40 pc at 1 Myr to 75 pc at an age of 1 Gyr, which is in agreement with the present estimate.

Increases in $z_h$ with the age can be explained in terms of the dynamical evolution of the Galactic disk. It is believed that the interactions of clusters with the disk and molecular clouds either evaporate the clusters or gradually disperse them from the GP with the time. Dynamical disruption is frequent for clusters that are closer to the center and mid-plane. The survivors reach large vertical distances from the plane. Therefore the scale height for the clusters in the Galactic disk gradually increases with age. The oldest clusters with ages of more than few gigayears are distributed nearly uniformly above the GP, hence they show very large scale heights. 
\subsubsection{Dependence of $z_h$ on $R_{GC}$}\label{ss_zh_Rgc}
\begin{figure}
\begin{center}
\includegraphics[width=8.5cm,height=5.5cm]{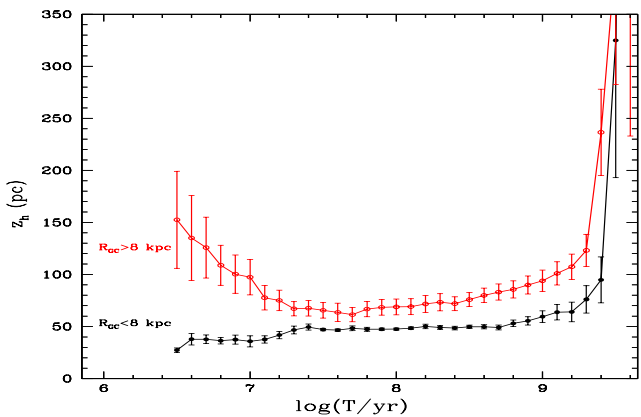}
\caption{Same as Fig.~\ref{Age_Zh}, but for two different $R_{GC}$ ranges.}
\label{Rgc_Rgc_Zh}
\end{center}
\end{figure}

Some recent studies found that the $z_h$ characterized by the clusters increases with $R_{GC}$ (e.g., Kent et al. 1991; Janes \& Phelps 1994; Bonatto et al. 2006), while some others found it to be independent of $R_{GC}$ (e.g., Buckner \& Froebrich 2014). To understand the correlation between these quantities in the present cluster sample, we built $z$-distribution in two different regions, that is, within the solar circle ($R_{GC}\leq8$) and outside the solar circle ($R_{GC}>8$). The respective number of clusters in these two regions is 621 and 597. Since any reliable determination of scale height needs a large enough sample of clusters to ensure good statistics, we determined $z_h$ in a $\Delta~logT = 1$ wide bin size to have enough clusters in each bin. However, to examine the progressive variation of $z_h$ with the age, we determined $z_h$ in overlapping intervals of $logT$ with a continuous increase of 0.1 in the $logT$. The resulting progressive variations of $z_h$ with $logT$ are shown in Fig.~\ref{Rgc_Rgc_Zh} for these two distributions in $R_{GC}$.

%

We estimate the average $z_h$ for $R_{GC}<=8$ kpc, and $R_{GC}>8$ kpc for the youngest population of clusters ($6 \leq log(T)<7$) to be $27\pm2$ pc (N=43) and $152\pm46$ pc (N=32), respectively. {Here, the number of clusters found in these regions are given in brackets. It should be noted that fewer clusters outside the solar orbit show a larger dispersion in $z,$ which has produced a larger uncertainty in the $z_h$ estimation. It is difficult to compare the cluster scale height with such a small sample of data, but while we found a two-times increase in $z_h$ between the inner and outer orbits for the older clusters, an increase of more than five times was found between the two orbits for the youngest populations of clusters. In a similar analysis containing a smaller sample of 654 clusters, Bonatto et al. (2006) also found an increase of a factor $\approx 2$ in $z_h$ between the
inside and outside of the solar orbit. In general, we found a gradual increase in $z_h$ with age for clusters older than $logT>7.3,$ and $z_h$ is smaller inside the solar orbit and larger outside the solar orbit in all age intervals. We also note that while YOCs show considerably different $z_h$ at two $R_{GC}$ values, the ratio of $z_h$ between the outer and inner orbits is almost constant for the older clusters except for the oldest sample of clusters, where $z_h$ is estimated with a large uncertainty.

It is well known that as a cluster becomes older, its chance of surviving decreases, and hence the number of clusters decreases with time. It is estimated that in the solar neighborhood, clusters with more than 6000 $M_\odot$ have a very low probability to survive beyond an age of 1 Gyr and only extremely massive clusters can survive beyond the 1 Gyr timescale. Buckner \& Froebrich (2014) furthermore speculated that older clusters might be dominated by some objects in their past that have undergone one or more violent interaction with the giant molecular clouds, which has moved them into orbits farther away from the GP where tidal forces
are weaker and collision rates relatively lower, hence their survival chances are higher. Those clusters that did not undergo such violent interactions stayed close to the GP and may not have survived to ages of a few Gyr.

The increase in scale height of the disk at larger $R_{GC}$ may be explained by the decrease in gravitational potential with increasing $R_{GC}$ , or in other words, by the weaker gravitational pull exerted by the disk or disk objects such as molecular clouds on the clusters as $R_{GC}$ increases. This may also explain why in the inner region of the Galaxy YOCs and IOCs are located mostly close to the GP: they experience strong gravitational pull from the disk or disk objects as well as the Galactic center. The difference in gravitational pull may arise from the decrease in density of the disk with the increasing $R_{GC}$.
\subsection{Solar displacement from the Galactic plane}\label{ss_z0}
The displacement of the Sun above the formal GP ($b = 0$ deg), which is denoted by $z_\odot$, has been inferred using different methods in the past. It is generally estimated by determining the mean vertical displacement of the symmetry plane defined by various objects such as the Cepheids, Wolf-Rayet objects, OB stars, and star clusters, and a wide range of $z_\odot$ has been reported (e.g., Joshi 2005, Bonatto et al. 2006, Reed 2006, Joshi 2007, Kong \& Zhu 2008, Buckner \& Froebrich 2014).  These studies give $z_\odot$ in the range of 6-30 pc depending upon different classes of objects (e.g., 15$\pm$3 pc, Conti \& Vacca 1990; 9.5$\pm$3.5 pc, Reed 1997, 2000; 27$\pm$4 pc, Chen et al. 2001; 16$\pm$5 pc, Elias et al. 2006; 6-28 pc, Joshi 2007, $18.5\pm1.2$ pc, Buckner \& Froebrich 2014). Considering the sample of YOCs and OB stars, Joshi (2007) concluded, however, that most of the disagreement in the estimation of $z_\odot$ comes from the difference in the determination method, data incompleteness, and selection of the data sample taken for different classes of astronomical objects. In the present study, we also noted that the  distribution of clusters is not symmetric around the GP, but is shifted toward
the southward disk, as shown in Fig.~\ref{z_hist}. The exponential fit in the figure gives a mean shift of $-6.2\pm1.1$ pc, which suggests that the mean vertical displacement of the Sun with respect to the GP based on the distribution of clusters is $z_\odot = 6.2\pm1.1$ pc. Bobylev \& Bajkova (2016) in a recent study
based on self-gravitating isothermal disk model obtained $z_\odot$ = 5.7$\pm$0.5 pc from the sample of 639 methanol masers, 7.6$\pm$0.4 pc from 878 HII regions, and 10.1$\pm$0.5 pc from 538 giant molecular clouds.
\subsection{Local mass density}\label{ldm}
It is noted that the progressive increase of the scale height begins only at an age of about 160~Myr. Before this, the evolution of the vertical distribution of clusters is determined by their ballistic motions in the gravitational field inside the thin disk, which have the form of harmonic oscillations about the GP with a characteristic period of $P_Z$~=~74~$\pm$~2~Myr (Dambis, 2003). This was first noted by Joeveer (1974), who found that the dispersion of vertical coordinates ($\sigma Z$) of Galactic classical Cepheids varies non-monotonically with age (inferred from the period-age relation) in a wave-like pattern. He attributed these oscillations to the fact that each sub-population of coeval Cepheids is not in vertical virial equilibrium, that is, the mean squared initial (vertical) velocities and coordinates of stars are not perfectly balanced. As a result, the stars at a certain $R_{GC}$, on an average, recede from and approach the GP in a correlated way with the period (frequency) determined by the local mass density (and therefore this is the same for all objects at a given $R_{GC}$). As a result, the mean squared displacement from the GP  varies periodically with twice the frequency (and half the period) of the vertical oscillations of stars about the GP. Joeveer (1974) used the inferred period of vertical oscillations to estimate the local mass density. We also performed the same analysis for the selected cluster sample. 

The frequency $\omega_z(R_G)$ of vertical oscillations about the GP is equal to
\begin{equation}
\omega_z(R_G)^2 = \left(\frac{\partial^2 \Phi}{\partial z^2}\right)_{R_G,0}\end{equation}
Here $\Phi$ is the Galactic potential (Binney \& Tremaine 2008) that obeys the Poisson equation:
\begin{equation}
4 \pi G \rho = \frac{1}{R} \frac{\partial}{\partial R}\left(\frac{\partial \Phi}{\partial R}\right) + \frac{\partial^2 \Phi}{\partial z^2},
\end{equation}
where $\rho$ is local mass density. The sum of the first two terms on the right-hand side can be expressed in terms of the local values of Oort's constants $A$ and $B$:
\begin{equation}
\frac{1}{R} \frac{\partial}{\partial R}\left(\frac{\partial \Phi}{\partial R}\right) = -2 (B^2 - A^2)
,\end{equation}
and hence
\begin{equation}
4 \pi G \rho = -2 (B^2 - A^2) + \omega_z(R_G)^2
\end{equation}
or
\begin{equation}
4 \pi G \rho = -2 (B^2 - A^2) + (2\pi/P_Z)^2
\label{density}
.\end{equation}
Based on the known values of Oort's constants and the period $P_Z$ of vertical oscillations, we can therefore determine the local mass density $\rho$. To estimate $P_Z$, we selected young clusters (t~$<$~160~Myr) from our sample located within $\pm$~1~kpc of the solar Galactocentric distance and $\pm$~200~pc of the GP, but excluded clusters located within 700~pc from the Sun to avoid the kinematical interference of the  Gould belt. To avoid the effect of eventual outliers, here we analyzed the variation of the median absolute deviation from the median $Z$ value, which we considered to be a much more robust estimator than dispersion $\sigma Z$. Figure~\ref{zdisp} shows the variation in the 20-point sliding median absolute deviation of the vertical coordinates of open clusters of our restricted sample and the corresponding periodogram, respectively. The periodogram reveals one conspicuous main period of $P$~=~42.8~$\pm$~0.7~Myr ($P_Z$~=~2$P$~=~95.6~$\pm$~1.4~Myr). We also performed a similar analysis for vertical velocity components whose median absolute deviation also varies periodically with half the period of the vertical oscillations. 
%
\begin{figure}
\begin{center}
\includegraphics[width=9.0cm,height=5.5cm]{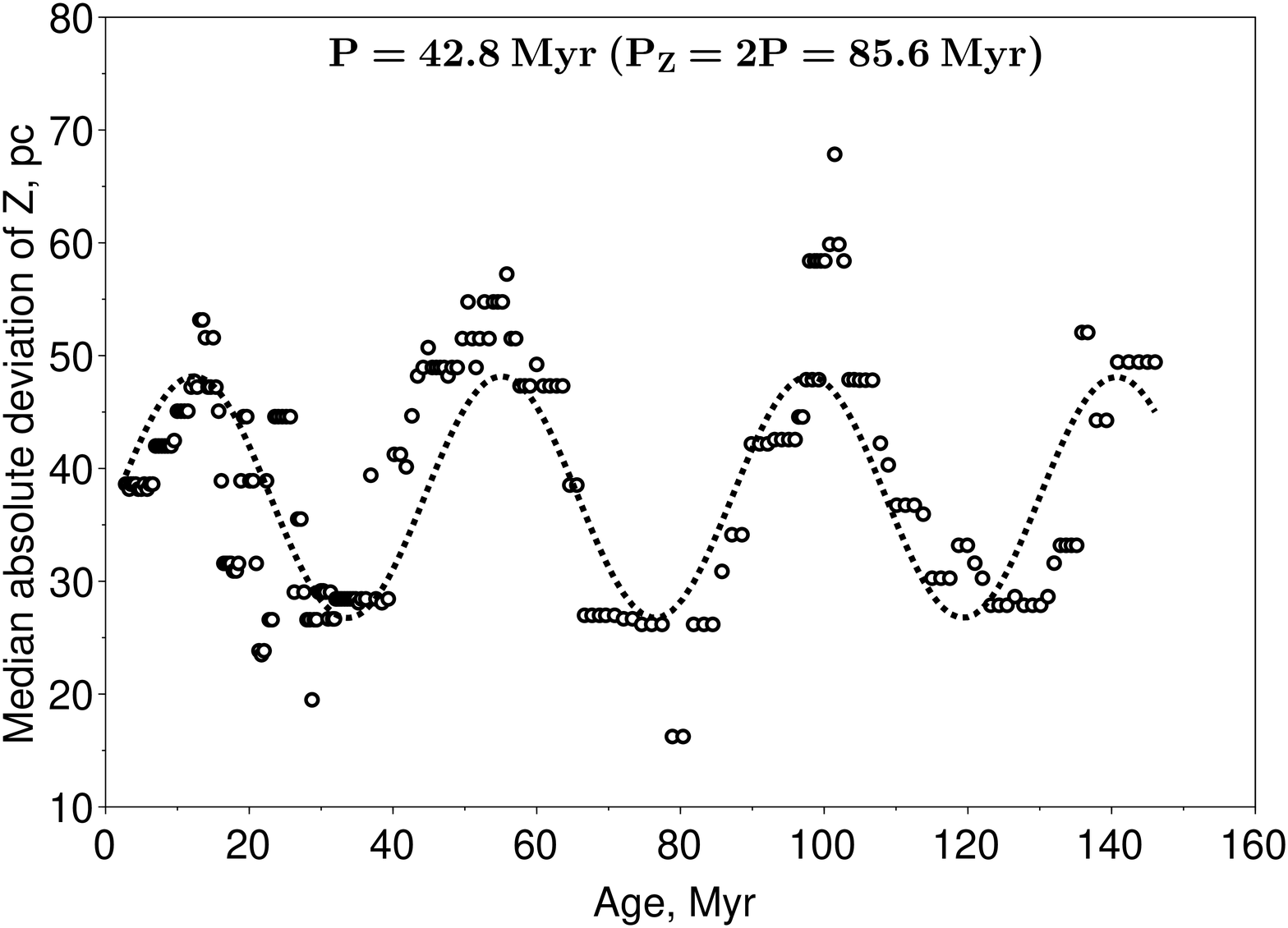}
\includegraphics[width=9.0cm,height=5.5cm]{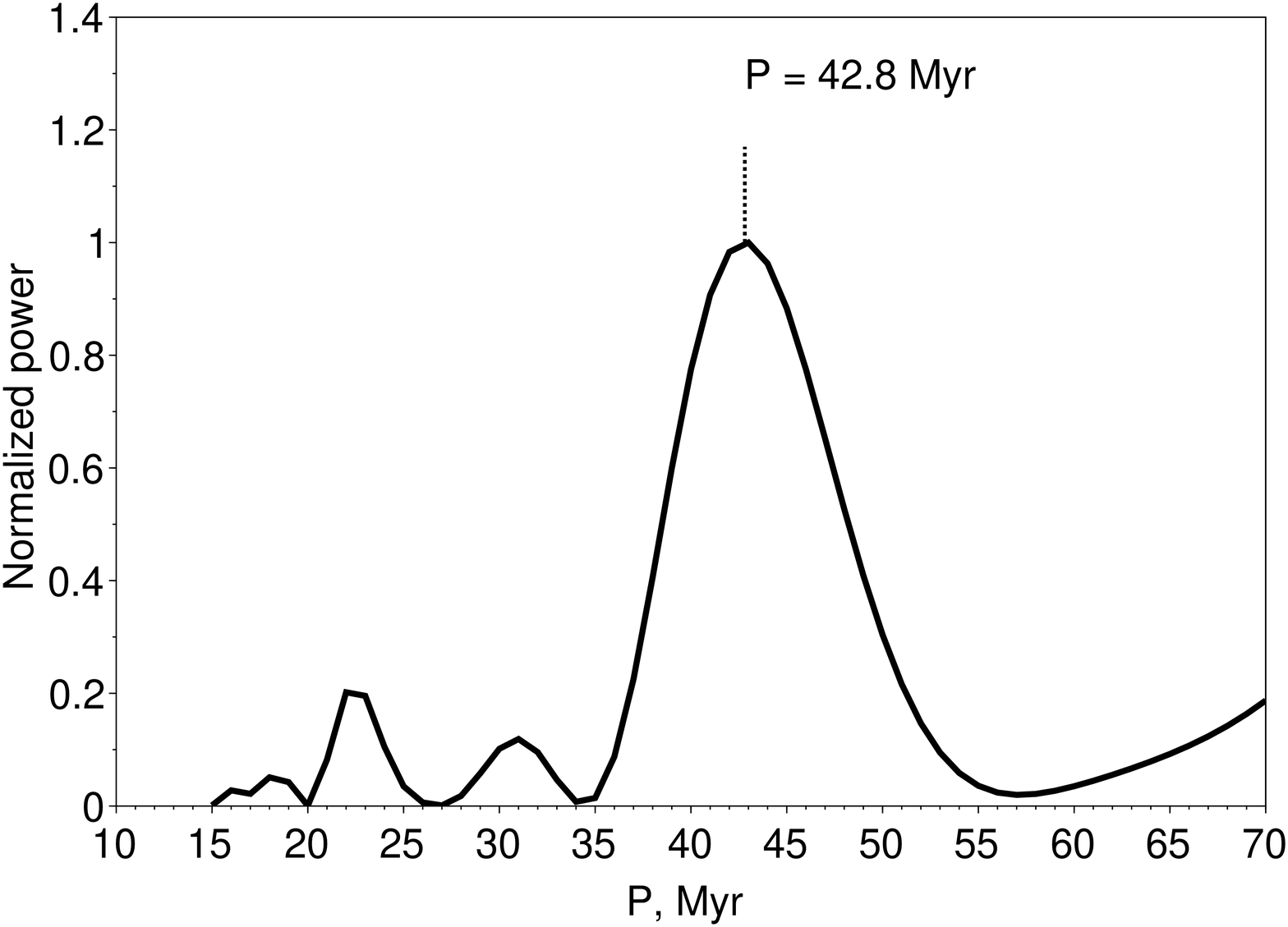}
\caption{Top panel: variation of the median absolute deviation of the vertical coordinates of young open clusters as a function of age. Bottom panel: periodogram  of this variation. The dotted line in the top panel shows the best-fit sinusoidal approximation of the absolute median deviation of $Z$ with the periodogram-peak period.}
\label{zdisp}
\end{center}
\end{figure}
%
\begin{figure}
\begin{center}
\includegraphics[width=9.0cm,height=5.5cm]{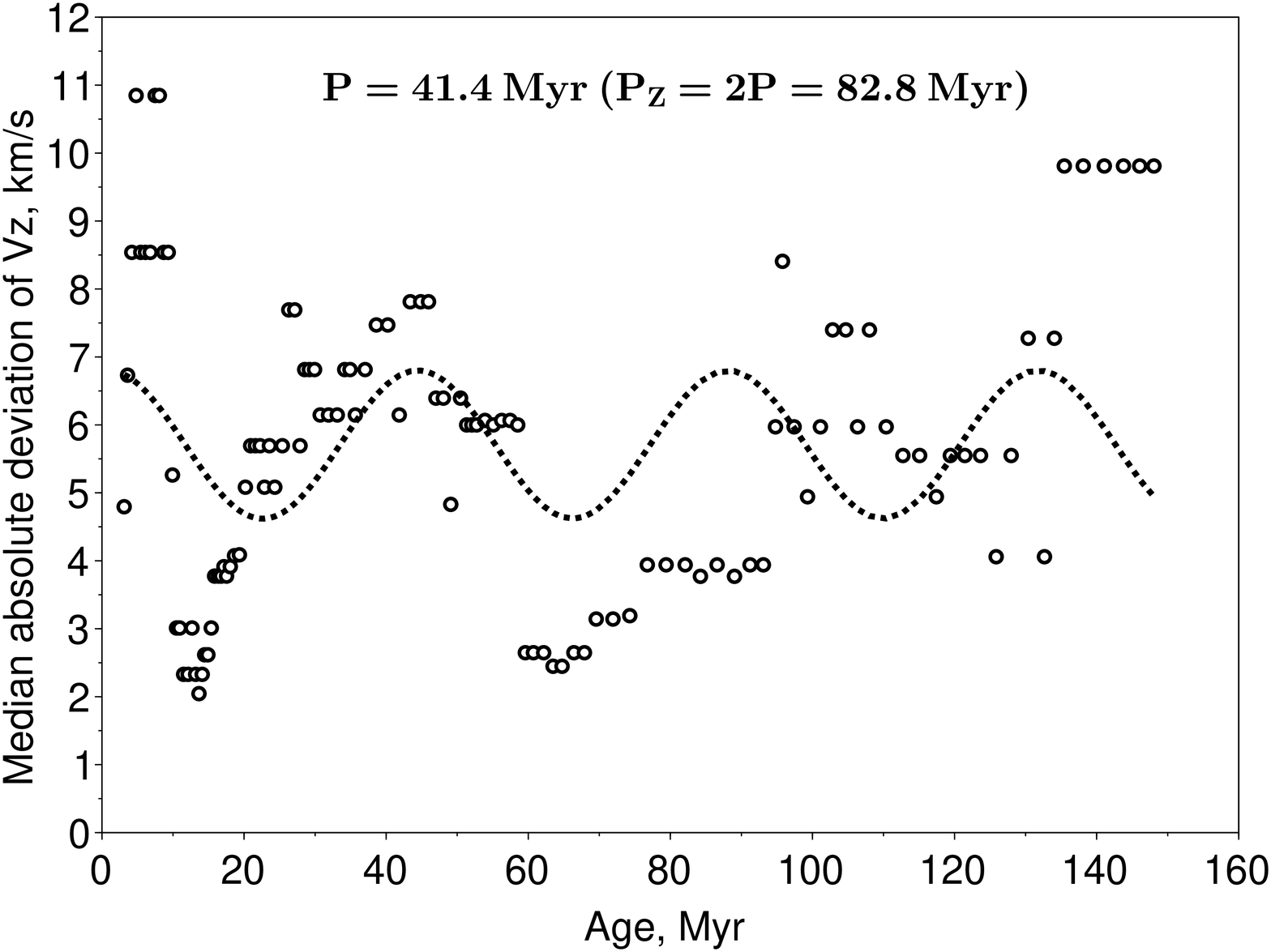}
\includegraphics[width=9.0cm,height=5.5cm]{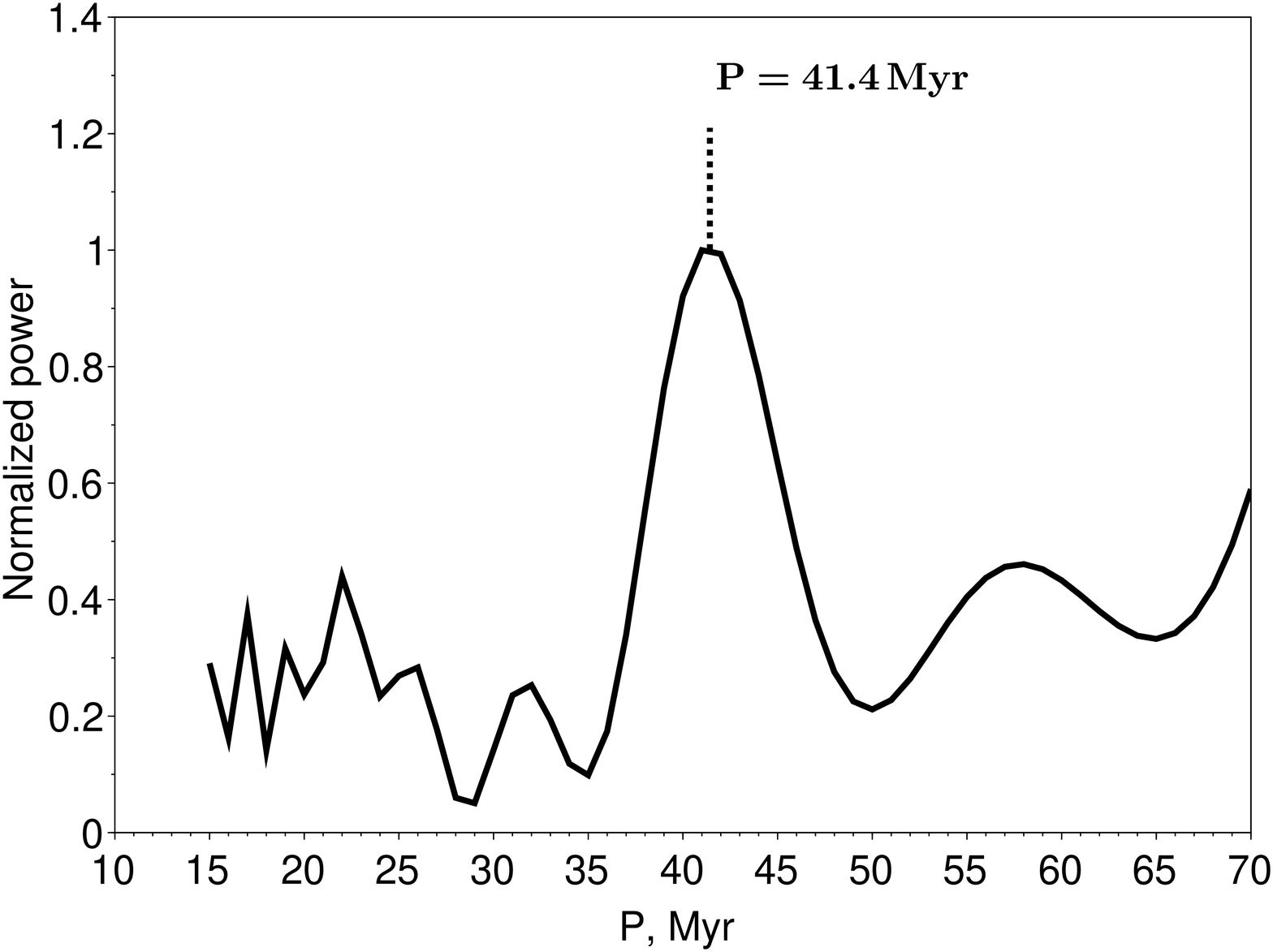}
\caption{Same as Fig.~\ref{zdisp}, but for the variation in the median absolute deviation of the vertical velocity components $\sigma~V_z$.}
\label{vzdisp}
\end{center}
\end{figure}
%
Figure~\ref{vzdisp} shows the variation in the 15-point sliding median absolute deviation of the vertical velocity components $\sigma Z$ of these open clusters  and the corresponding periodogram, respectively. Again, we see a peak at $P$~=~41.4~$\pm$~1.4~Myr ($P_Z$~=~2$P$~=~82.8~$\pm$~2.8~Myr). We note, however, that the peak in this periodogram is less conspicuous than for the variations in the median coordinate deviation, and the corresponding sine-curve fit approximates the actual variation rather poorly.

Using these period estimates in Eq.~\ref{density}, we find two local mass density estimates: 0.088~$\pm$~0.004~$M_{\odot}/{\rm pc}^3$ and 0.094~$\pm$~0.006~$M_{\odot}/{\rm pc}^3$. Here we have taken Oort's constants A = 14.8 km/s/kpc and  B = -13.0 km/s/kpc from Chen et al. (2008). The estimated periods and inferred local mass densities both agree well with each other, but are somewhat lower than earlier results based on the analysis of the open cluster catalog of Dambis (1999) for standard stellar evolution models of Pols (1998) without  overshooting: $P_Z/2$~=~37~$\pm$~1~Myr ($P_Z$~=~74~$\pm$~2~Myr) and 0.118~$\pm$~0.006~$M_{\odot}/{\rm pc}^3$, respectively. Our result is marginally consistent with the local mass density estimates based on Hipparcos parallaxes and proper motions combined with radial-velocity data for A- and F-type stars, $\rho_0$~=~0.102~$\pm$~0.010~$M_{\odot}/{\rm pc}^3$ (Holmberg \& Flynn, 2000) and bright red giants, $\rho_0$~=~0.100~$\pm$~0.005~$M_{\odot}/{\rm pc}^3$ (Korchagin et al. 2003). Given the local volume density estimate for visible matter, $\rho_{vis}$~=~0.095~$M_{\odot}/{\rm pc}^3$ (Holmberg \& Flynn, 2000), our result implies the absence of or very small amount of dark matter in the nearest solar neighborhood in the Galactic disk, which is consistent with the most recent estimates, for example, $\rho_{dm}$~=~0.018~$\pm$~0.005~$M_{\odot}/{\rm pc}^3$ (Xia et al. 2016, based on LAMOST data for G- and K-type MS stars), $\rho_{dm}$~=~0.009~$M_{\odot}/{\rm pc}^3$ (McGaugh 2016, from the terminal velocity curve and the mass discrepancy-acceleration relation), and $\rho_{dm}$~=~0.013~$\pm$~0.003~$M_{\odot}/{\rm pc}^3$ (McKee et al., 2015, reanalysis of earlier data).
\subsection{Surface mass density}\label{smd}
We now use our estimated scale height to determine the surface density $\Sigma$ of all the matter present in the Galactic disk. The surface mass density  $\Sigma_{out}(z_{out})$ within $\pm z_{out}$ ($z_{out}$ is the half-thickness of the disk layer inside which the surface density is computed) of the GP is equal to (see Korchagin et al.~2003)
\begin{equation}
 \Sigma_{out}(z_{out}) = - {\overline {v_z^2} \over  2\pi G }
 \Big( {1 \over \rho_i} {\partial \rho_i \over \partial z} \Big)\Big{\vert}_{z_{out}}
   + { 2 z_{out} (B^2 -A^2) \over 2\pi G}
   \label{eq1}
,\end{equation}
where the logarithmic derivative $\Big( {1 \over \rho_i} {\partial \rho_i \over \partial z} \Big) $~=~$1 \over z_h$ is equal to the inverse scale height; $A$ and $B$ are Oort's constants of Galactic rotation, which are accurately known, and $\overline {v_z^2}$ is the vertical velocity dispersion of the cluster sample considered. The latter quantity is very difficult to estimate for clusters because most of them lie close to the GP and therefore their computed vertical velocity components $v_z = V_R sin(b) + d \mu_b cos(b)$ are dominated by the proper-motion term $d \mu_b cos(b)$ and hence, the errors of the latter,
$$
\sigma {(d \mu_b cos(b))^2} =  (\sigma d)^2 \mu_b cos(b)+ (d \sigma \mu_b)^2 cos(b)
$$
(which, in turn, are dominated by those of the $\mu_b$ proper-motion component), become comparable to or exceed substantially the actual vertical velocity dispersion even  at not too large heliocentric distances (about 1--2~kpc). Here we adopted the $\overline {v_z^2}$~=~8.0~$\pm$~1.3~km/s estimate determined by Chen et al. (2008) for 117 clusters younger than 0.8~Gyr and located in the 0.5--2.0~kpc heliocentric distance interval. Our scale height is equal to $z_h$=60~$\pm$~2~pc for the cluster sample selected using the same age and distance constraints. By combining it with the above $\overline {v_z^2}$ estimate  and Oort's constant values $A$~=~14.8~km/s/kpc and $B$ = -13.0~km/s/kpc, we obtained a Galactic-disk surface density of $\Sigma$~=~40~$\pm$~12~$M_{\odot}/pc^2$. Our estimate agrees well with the surface density of $\Sigma$~=~42~$\pm$~6~$M_{\odot}/pc^2$ within $\pm$~350~pc of the GP based on a sample of 1476 old bright red giants from the {\it Hipparcos} catalog (Korchagin et al. 2003).
\section{Correlation between various parameters}\label{s_correlation}
The correlations between various cluster parameters provide important constraints to our understanding of the processes of cluster formation and dynamical evolution. With such a large data set available in the present study, it is interesting to investigate the correlations among different cluster parameters, particularly those that reflect the formation mechanism of the star clusters.
%
\begin{figure}
\includegraphics[width=9.0cm,height=5.5cm]{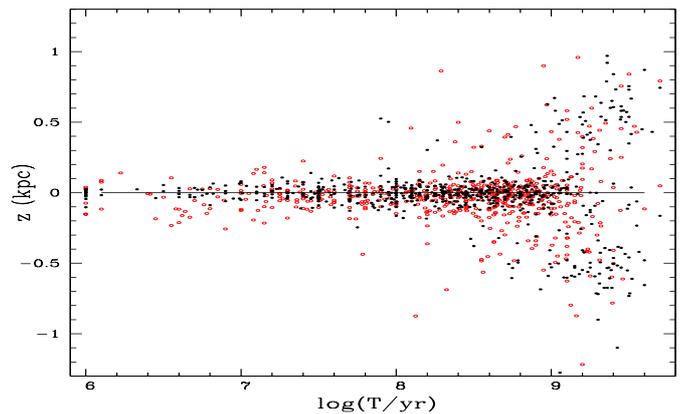}
\caption{Distribution of clusters in the z-axis as a function of age. The filled and open circles represent clusters inside and outside of the solar orbit, respectively.}
\label{age_z}
\end{figure}
%
\subsection{Age vs $z$}\label{ss_age_z}
In Fig.~\ref{age_z} we plot $z$ against $logT$ , which shows a torch-like structure. The clusters that lie within the solar orbit are shown by filled circles, while those found outside the solar orbit are shown by open circles. It is quite conspicuous from the figure that the clusters are more confined close to the GP in the direction of Galactic center, but they are scattered and tend to leave the GP in the anticenter direction. Almost all the YOCs lie close to the GP, and their distribution has an approximately constant width of $z \approx 60$ pc in the vertical direction. It is known that clusters younger  than about 20 Myr remain embedded within the molecular clouds in their parent arm (Dias \& L{\'e}pine 2005, Moitinho 2010). As the embedded cluster evolves, it loses mass, and in the process the cluster becomes disrupted and dissolves into the field. It is believed that only about 5\% of the clusters survive to progress further (Lada \& Lada 2003), and at around 20 Myr, they drift away from the arm and join the general disk population (Moitinho 2010). As time progresses, clusters  move farther away from the Galactic disk as a result of several factors, including tidal interactions with the Galactic disk and the bulge and mass loss due to stellar evolution. Finally, only a few clusters survive to reach the outer disk of the Galaxy. The OOCs that have moved to the outer disk are now less prone to disruptive forces, hence have a higher chance of surviving.

The cluster distribution along the GP for different age bins indicates that young and intermediate-age clusters are not only very close to the GP, but are also uniformly distributed around the GP. However, most clusters that are older than 1 Gyr are found in the outer disk. It is believed that the disk evolves  with time, that is, thickens with age. This might be a combined effect of many factors such as selective destruction of clusters at lower Galactic heights by the Galactic tidal field, spiral arm shocks, stellar evolution, and encounters with the molecular clouds (Lamers \& Gieles 2006, Moitinho 2010).
\subsection{Reddening vs $z$}\label{ss_ebv}
Star clusters are excellent tools for mapping the Galactic extinction. As most of the clusters reside in the Galactic disk, they are subject to strong and irregular interstellar extinction depending upon their positions in the Galaxy. The $E(B-V)$ ranges from negligibly low value for nearby clusters or objects at high Galactic latitude to values as high as $\sim$ 3.7 mag for highly obscured star-forming regions near the GP. Over half of the observed clusters have an extinction of $E(B-V) \lesssim 0.3$ mag, and only a few clusters are embedded in regions of high extinction.

In Fig.~\ref{z_ebv} we draw the mean $E(B-V)$ distribution as a function of mean $z$. To estimate the mean $E(B-V)$, we made a bin of $z$ = 50 pc. As expected, the reddening is found to be strongest close to the GP and to decrease gradually outwards in the directions of both northern and southern disk due to the
decrease in the amount of dust and gas. Taking a sample of 722 star clusters, Joshi (2005) found that approximately 90\% of the reddening material lies within a distance of $430\pm60$ pc of the GP. We noted that at any vertical position from the GP, a range of cluster reddening is found. This is well understood as clusters are observed at different longitudes for the same vertical position from the GP, and reddening material is not uniformly distributed in the Galaxy. Reddening is comparatively higher and variable toward the Galactic center and systematically decreases toward the anticenter direction (Joshi 2005). We drew separate least-squares linear fits for the $z$-$E(B-V)$ distributions both above and below the GP, as shown by the continuous lines in Fig.~\ref{z_ebv}, and found the following slopes
$$dE(B-V)/dz = 0.40\pm0.04~mag/kpc ~~~~~~(z~\leq~0)$$
$$dE(B-V)/dz = 0.42\pm0.05~mag/kpc ~~~~~~(z~\geq~0).$$
This implies that reddening varies in a similar manner in the
two directions of the Galactic mid-plane. As most of the younger clusters lie within the GP, it also suggests a general trend in the decrease of reddening with age, that is, younger clusters, which are located in the spiral arms of the Galaxy, tend to be more reddened than the older ones.
%
\begin{figure}
\includegraphics[width=9.0cm,height=5.5cm]{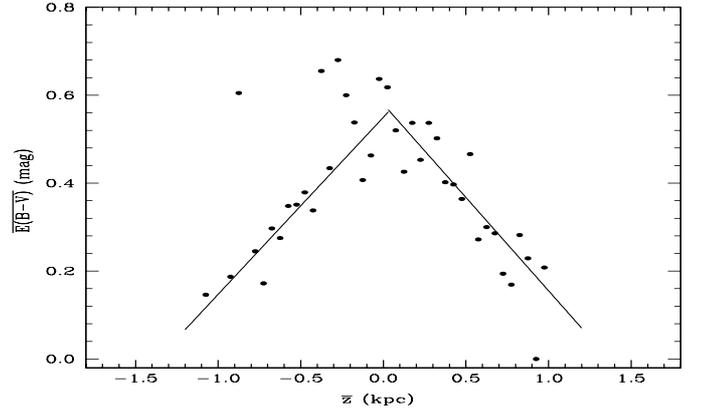}
\caption{Mean E(B-V)  as a function of mean $z$. The two separate lines represent the least-squares linear fits in $z<0$ and $z>0$ regions.}
\label{z_ebv}
\end{figure}
%
\subsection{Mass-radius relation}\label{ss_mass_dia}
The mass-radius relation for the clusters is very important for understanding the dynamical evolution of cluster populations. The reason is that during their early evolution phase, clusters suffer from numerous encounters with the massive molecular gas clouds from which they form and in the process loose both mass and size. The timescale of the cluster disruption during such encounters depends upon their densities (Spitzer 1958). The mass-radius relation thus determines the properties of the clusters that survive such encounters with the massive molecular clouds (Gieles et al. 2006, Elmegreen 2010). While several studies found no direct correlation between mass and radius for the young massive clusters of different galaxies (Bastian et al. 2005, Scheepmaker et al. 2007,  Portegies et al. 2010),  others have noted a slight positive mass-radius correlation for massive objects ($> 10^6~M_\odot$) such as globular clusters and bright elliptical galaxies (e.g., Zepf et al. 1999, Hasegan et al. 2005, Kissler-Patig et al. 2006).

It is well known that molecular clouds follow a mass-radius relation, but it is still not clear whether this is true for the open clusters, which are formed after the collapse and fragmentation of the same molecular clouds. However, to study any correlation between them, it is necessary to have precise measurements of the mass and age of the clusters. While cluster ages can be determined with some precision, there is no direct way to measure the total masses of clusters without exact knowledge of individual membership detail of the stars found in the cluster field, and hence they remain one of the most elusive fundamental parameters of star clusters to be determined accurately. Piskunov et al. (2007) have indirectly inferred the tidal masses for the 236 clusters in the solar neighborhood based on the cluster tidal radii, which they have determined by fitting a three-parameter King's profile (King 1962) to the observed integrated density distribution. Furthermore,  Piskunov et al. (2008) determined masses of 650 clusters, including the previous 236 clusters, from a relation between the semi-major axis of the density distributions of cluster members and the fitted King radius. Their study does not give an exact estimate of the total mass of the clusters, but they provide approximate values for the clusters masses that can be used to examine whether less massive ($< 10^6~M_\odot$) star clusters also exhibit a similar mass-radius relation as their massive counterparts.
%
\begin{figure}
\includegraphics[width=9.0cm,height=8.7cm]{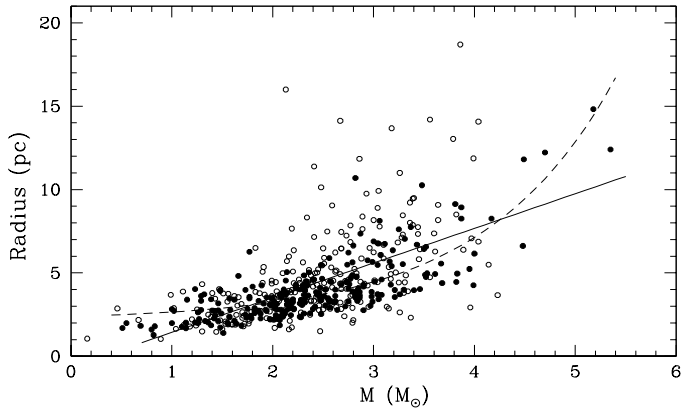}
\vspace{-3.6cm}
\caption{Mass-radius distribution for the clusters. The filled and open circles represent clusters within and outside the solar orbit. The continuous line represents the best-fit linear regression line. The dashed curve shows the $R \propto M^{1/3}$ relation for the clusters within the solar orbit assuming a constant cluster density.}
\label{dia_mass}
\end{figure}
%

To examine the correlation between the observed distribution of size and mass of the clusters, we plot the cluster radius versus total mass in logarithmic scale in Fig.~\ref{dia_mass}. Not many massive clusters with smaller radius are present in the mass-radius diagram in Fig.~\ref{dia_mass}. There is a possibility that the massive clusters with small diameters are not able to survive longer as they dissolve as a result of the numerous encounters among cluster members. Similarly, a lack of low-mass clusters with large diameter is also seen in the diagram because such clusters would not be able to survive for too long as they become unstable in the Galactic tidal field and other disruptive forces. In Fig.~\ref{dia_mass} the scatter is largest for the massive clusters, but in general, it shows a positive correlation between the two parameters. We obtained the following linear relation from the best fit between mass and radius of the cluster sample:
\begin{equation}
R = 2.08(\pm0.10)~log(M/M_\odot) - 0.64(\pm0.27)
.\end{equation}
The continuous line superimposed on the data represents the best-fit linear curve in Fig.~\ref{dia_mass}. Since the correlation between mass and radius has some inference on the cluster density, which is directly proportional to the cluster lifetime, we also fit a power-law function $R = a~M^{1/3} + b$ for a constant cluster density. However, the best-fit power law, shown by the dashed line in Fig.~\ref{dia_mass}, gives a reduced chi-square of 3.66 per degree of freedom (NDF=490), which is slightly lower than
the 3.58 found for a linear fit. We therefore conclude that a linear relation is slightly better suited to the cluster mass-radius relation. 

When we separately drew clusters inside and outside the solar orbit in Fig.~\ref{dia_mass}, we found that clusters within the solar orbit show a more systematic variation in the mass-radius diagram than clusters outside the solar orbit. It is likely that the massive clusters with larger radius have a lower chance of surviving the external shocks in the direction of the Galactic center, hence most such clusters are already disrupted and dissolved into the field inside the solar orbit. When we drew a power-law function $R = a~M^{1/3} + b$ for only those clusters that lie within solar orbit, we found that it fits better with a reduced chi-square of 1.39 per degree of freedom (NDF=255) than the linear fit, which yields a reduced chi-square of 1.61 per degree of freedom. A more clear inference can, however, be drawn from the mass-radius diagram only when we have more accurate cluster masses.
\subsection{Mass-age relation}\label{ss_mass_age}
We know that star clusters are destroyed over time through stellar and dynamical evolution, and by the external forces of interactions with the Galactic tidal field or with the molecular clouds. The dissolution times of the clusters depend on several factors such as the initial mass of the cluster, orbital location, and internal structure. More massive clusters are believed to survive longer than low-mass clusters, and \textcolor[rgb]{1,0.501961,0}{\textcolor[rgb]{0,0,0}{poor }}}clusters disrupt easier than massive ones (Tadross et. al. 2002, Lamers \& Gieles 2006). In the process of destruction, they undergo a significant amount of mass loss in their lifetimes. To study the evolution of cluster masses in our data set, we drew the total mass of the clusters against their age in the $logM-logT$ plane. Figure~\ref{age_mass} shows the variation in $logM$ with $logT$ for 489 clusters with available mass estimates. The dark points are the mean of all the points for an age bin of $\Delta logT = 0.5$ and the error bars represent the variance about the mean. The more massive clusters are among the younger ones, and there is some correlation between $logM$ and $logT$ of the clusters. A least chi-square ($\chi^2$) straight line fit yields the following relation:
\begin{equation}
log(M/M_\odot) = -0.36(\pm0.05)~logT + 5.3(\pm0.4)
,\end{equation}
where $M$ is the total mass of the cluster in units of solar mass and $T$ in yr. The Pearson correlation coefficient was found to be -0.95, which shows a strong correlation between the two parameters. It can be concluded from Fig.~\ref{age_mass} that the younger clusters have higher masses and there is a decreasing trend in the total masses of the clusters with the age. This is because the younger clusters contain relatively more stars of higher mass. 

%
\begin{figure}
\includegraphics[width=9.0cm,height=5.5cm]{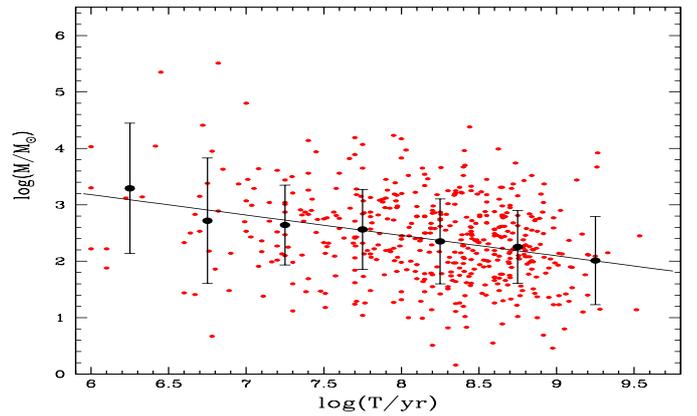}
\caption{log-log relation between age and total mass of the clusters. The solid points are the mean value of log M in the interval of $\Delta logT=0.5$, while the error bars are the variance about the mean. A least-squares linear fit is also drawn as a continuous line.}
\label{age_mass}
\end{figure}

The mean $log(M/M_\odot)$ for YOCs, IOCs, and OOCs is 4.07, 2.99, and 2.71. The corresponding sub-samples contain 61, 371, and 57 objects, respectively. For the YOCs, the mean cluster mass is highest. This suggests that while the youngest population of clusters contains the highest mass, the mass-loss rate is also highest for them. For 1-10 Myr age clusters, the typical mass-loss rate during their evolution is about 150 $M_\odot$ Myr$^{-1}$. With time, the total cluster mass and the mass-loss rate decrease substantially to about 0.05 $M_\odot$ Myr$^{-1}$ for clusters older than 1 Gyr. This confirms that the mass-loss rate $dM/dT$ increases with cluster mass but decreases with age. The fundamental reason of the mass loss in clusters is basically dictated by two factors, one is mass loss due to stellar evolution of massive stars, and the other is the dynamical evolution that
primarily affects the low-mass stars. It is well known that in the earliest phase of the cluster evolution, mass loss primarily occurs from the stellar evolution of the most massive stars in the cluster (R\"{o}ser et al. 2010). Sometimes the ejection of massive stars could also reduce the cluster mass in YOCs. For example, when Schilbach \& R\"{o}ser (2008) traced back the trajectories of so-called field O stars, they found that most of them originated in a younger population of clusters. In the youngest clusters, the total mass loss is less significant because at this age, only very high-mass stars are stripped off from the clusters, and these massive stars ($>30 M_\odot$) do not contribute significantly to the total mass of the cluster (Lamers \& Gieles 2006). As time progresses, the low-mass clusters disrupt completely, while high-mass stars in the clusters are entirely evolved. This reduces the total mass in the case of IOCs. When the clusters become older, only low-mass stars are left in them, hence they are the least massive of all the populations.

The encounters between cluster members combined with the Galactic tidal field also cause the slow and gradual evaporation of the cluster members into the field, preferentially low-mass ones, hence reduce the cluster mass over the time. A close encounter and stellar evolution can additionally speed up this process, especially for small, sparsely populated clusters (de la Fuente Marcos \& de la Fuente Marcos 2004). Small poorly populated clusters dissolve primarily as a result of the internal dynamical effects over time on the order of a few hundred million years. For richer massive clusters the Galactic tidal effects become significantly moderated by the cluster mass. In the case of OOCs, while clusters
with low initial mass have ceased their life, more massive clusters still survive despite loosing a significant amount of their mass in the cluster evolutionary process. Nonetheless, the presence of high masses in a few OOCs can also be noted in Fig.~\ref{age_mass}, which could be attributed to either a very high initial mass of the clusters, a different dynamical state of clusters or to the fact that they may be located in regions where disruptive forces may not be very active. In many past studies, a mass-age relation has been noticed for clusters, not only in our Galaxy (Pandey et al. 1987) but in the Large and Small Magellanic Clouds (Hunter et al. 2003; de Grijs \& Anders 2006; Chandar et al. 2010), dwarf galaxies such as IC 1613, DDO 50, NGC 2366 (Melena et al. 2009), and spiral galaxies (Mora et al. 2009). A power-law correlation was also found by Baumgardt \& Makino (2003) from the N-body simulation and by Boutloukos \& Lamers (2003) using clusters taken from four different galaxies of the Local Group.
\subsection{Cluster lifetime vs diameter}\label{lifetime}
The lifetime of a cluster is estimated as the period between its formation and disintegration. Theory predicts that the lifetime of clusters depends on their initial mass, and rich clusters have longer lifetimes than their poorer counterparts (Gnedin \& Ostriker 1997, Baumgardt \& Makino 2003). A similar conclusion was also made by Tanikawa \& Fukushige (2005), Carraro et al. (2005) and Bonatto et al. (2012), who suggested that poorly populated clusters do not survive longer than a few hundred Myr, whereas rich ones survive for a longer period. Based on their analysis of 125 clusters with fewer than 100 stellar members, Pandey \& Mahra (1986, 1987) concluded that lifetimes of clusters not only depend on the richness of the clusters, but also on their sizes. Since the observed age distribution obtained in their study was based on a significantly small sample of clusters, here we study the longevity of the clusters on the basis of a larger sample of clusters using the MWSC catalog where the most probable members (MPMs) are given for the listed clusters. We prepared two samples of clusters, one containing poor clusters with fewer than 100 MPMs, and the other containing rich clusters with more than 100 MPMs. To avoid including the stellar associations, we did not include clusters younger than 10 Myr in our analysis. For both sets of clusters, we determined the cluster frequency $\nu(\tau)$ = $\Delta n/\Delta \tau$ as a function of age (in $10^8$ yr) for different ranges of diameters. The age distributions of clusters for various linear diameter intervals were then calculated from the observed frequency of the clusters. The percentage of clusters $P(\tau)$ that reached the age $\tau$ was obtained using the relation 
$$
P(\tau) = \nu(\tau)/\nu(0),
$$ 
where $\nu(0)$ is the initial frequency of the clusters.

\begin{figure}
\includegraphics[width=9.0cm,height=10.5cm]{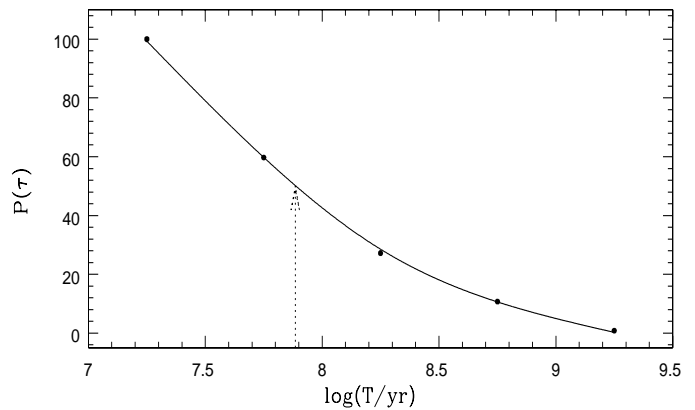}
\vspace{-5cm}
\caption{The variation of P($\tau$) for the different age intervals. The solid line shows best cubic line fit while the dotted line indicates the $t_{/12}$ position along the x-axis (see text for detail).}
\label{P_tau}
\end{figure}
%
\begin{figure}
\includegraphics[width=9.0cm,height=10.5cm]{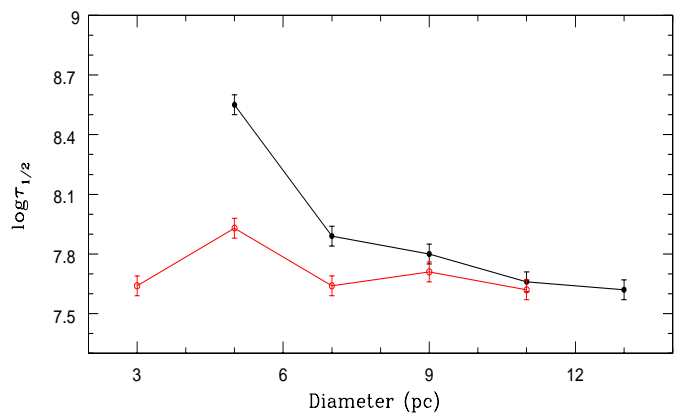}
\vspace{-5cm}
\caption{The variation of lifetime of clusters ($\tau_{1/2}$) as a function of linear diameter for the clusters having $N<100$ (open circles) and $N\geq100$ (filled circles).}
\label{dia_tau}
\end{figure}
%

The total lifetime $\tau_{1/2}$ of the clusters was obtained statistically from the observed age distribution $\nu(\tau)$ and defined as the time when 50\% of the total clusters have disintegrated in a given sample. A sample plot is illustrated in Fig.~\ref{P_tau} to show the $\tau_{1/2}$ estimation in the diameter interval 6-8 pc for the rich clusters. The distribution of $\tau_{1/2}$ of the clusters as a function of their linear diameters is shown in Fig.~\ref{dia_tau} for the two richness classes. Here we assigned a constant error of 0.05 in the estimation of $\tau_{1/2}$ based on the combined error in the age determination and the fit of the slopes. Assuming that the formation rate is constant, the variation of $\tau_{1/2}$ over diameter suggests that the lifetimes of rich clusters decrease for the larger-sized clusters, but poor clusters have an almost uniform disintegration time regardless of their sizes. These trends also imply that clusters with higher stellar density have longer lifetimes than lower density clusters of the same size. This is expected since poorly populated clusters are more easily destroyed by the interactions with the Galactic disk, the tidal pull of the Galactic bulge, frequent collisions with the giant molecular clouds and other such factors, while rich clusters can survive these external shocks for a longer time.
\section{Concluding remarks}\label{conclu}
The properties of Galactic star clusters and the observed correlations between their various parameters offer important empirical constraints not only on the cluster formation, but also on the history of the Galaxy. However, any meaningful study of star clusters and their role to understand Galactic structure is very much dependent on the accurate determination of the cluster parameters and the
degree of completeness of the cluster sample. Recently, the MWSC project has provided the parameters of more than 3000 clusters that were determined in a very homogeneous manner. To understand the Galactic structure from an observational perspective, we have statistically analyzed their data to obtain the distribution and correlations between various parameters using an almost complete sample of clusters within 1.8 kpc. Our main results and conclusions are summarized as follows:
\begin{itemize}
\item The dependence of the spatial distribution of clusters on age indicates a complex interplay between cluster formation and survival within the Galaxy. An analysis of the distribution of clusters in different longitudes has revealed many clusters in active star-forming regions such as the Carina arm, whereas the deepest minima are observed in the region of Sagittarius.
\item  We estimated an average Galactic disk scale height as $z_h = 60\pm2$ pc for clusters younger than 700 Myr. However, this increases to $64\pm2$ when we include old age clusters.
\item We found that the scale height is strongly dependent on $R_{GC}$ and age. On an average, the scale height is more than twice larger in the outer region than in the inner region of the solar circle, except for the youngest clusters, where it is as much as five times high.
\item The distribution of clusters in the Galaxy is shifted slightly below the GP, and we estimated the solar vertical displacement to be $z_\odot = 6.2\pm1.1$ pc in the southward direction.
\item The periodic age dependence of the mean squared displacement of young local clusters from the GP implies a local mass density of $\rho_0$~=~0.090~$\pm$~0.005~$M_{\odot}/{\rm pc}^3$. We found a negligibly small amount of dark matter in the Galactic disk around the solar neighborhood.
\item We found a strong correlation between the extinction of the cluster and its position in the GP with respective slopes of $dE(B-V)/dz = 0.40\pm0.04$ and $0.42\pm0.05$ mag/kpc below and above the GP.
\item There is a correlation between the mass and linear diameter of clusters, with massive clusters being larger in size.
\item It is found from the distribution of the total mass of the clusters as a function of age that the younger clusters have higher masses, and there is a decreasing trend in the total masses of the cluster with age, suggesting that the clusters erode over time in a regular manner. A linear gradient of $-0.36\pm0.05$ was estimated between mass and age in the log-log scale.
\item The lifetimes of rich clusters are found to be decreasing with the increase in the sizes of the clusters, and denser clusters survive for longer than the lower density clusters of the same size. 
\end{itemize}

The geometrical and physical characteristics of a statistically significant number of clusters have allowed us to understand the large-scale spatial properties of the cluster systems in the Galaxy. The structural and physical parameters of clusters enabled us to check the mutual correlations between the individual parameters. The homogeneous nature and large size of the sample of cluster data leads us to conclude that the correlations we
derived display real trends. One of the significant results is that the properties of clusters in the Galactic disk change for the cluster samples of different ages. We would like to mention here that even though much progress has been made in understanding the general properties of open star clusters and the Galactic structure in recent years, the continuous increase and upgrading of the cluster samples and cluster parameters to be provided by forthcoming all-sky surveys such as GAIA and their accurate mass estimation would help us immensely to further strengthen our knowledge of these systems and to understand our Galaxy in a better way.
\section*{Acknowledgments}
We wish to thank referee for useful comments that improved the clarity of this paper. We are grateful to Rohit Nagori, who contributed in the initial stage of the present work. The work presented here has been carried out under the Indo-Russian project DST/INT/RFBR/P-219 (Russian Foundation for Basic Research, grant no. 15-52-45121). AKD acknowledges the support from the Russian Foundation for Basic Research (grant no. 14-02- 00472a). Analysis of the vertical oscillations of the cluster layer and estimation of the local mass density was supported by the Russian Scientific Foundation grant no. 14-22-00041.
\end{document}